\documentclass[10pt,aps,pre,twocolumn,superscriptaddress,floatfix,longbibliography]{revtex4-1}
\usepackage{graphicx}
\usepackage{amsmath}
\usepackage{amssymb}
\usepackage{hyperref}

\begin{document}
\title[Anisotropy in electrical conductivity]
{Anisotropy in electrical conductivity of two-dimensional films containing aligned rodlike  particles: continuous and lattice models}

\author{Yuri Yu. Tarasevich}
\email[Corresponding author: ]{tarasevich@asu.edu.ru}
\affiliation{Laboratory of Mathematical Modeling, Astrakhan State University, Astrakhan, Russia, 414056}

\author{Nikolai I. Lebovka}
\email{lebovka@gmail.com}
\affiliation{Department of Physical Chemistry of Disperse Minerals, F. D. Ovcharenko Institute of Biocolloidal Chemistry, NAS of Ukraine, Kiev, Ukraine, 03142}
\affiliation{Department of Physics, Taras Shevchenko Kiev National University, Kiev, Ukraine, 01033}

\author{Irina V. Vodolazskaya}
\email{vodolazskaya_agu@mail.ru}
\affiliation{Laboratory of Mathematical Modeling, Astrakhan State University, Astrakhan, Russia, 414056}

\author{Andrei V. Eserkepov}
\email{dantealigjery49@gmail.com}
\affiliation{Laboratory of Mathematical Modeling, Astrakhan State University, Astrakhan, Russia, 414056}

\author{Valeria A. Goltseva}
\email{valeria.lozunova@gmail.com}
\affiliation{Laboratory of Mathematical Modeling, Astrakhan State University, Astrakhan, Russia, 414056}

\author{Valentina V. Chirkova}
\email{valyffkin@mail.ru}
\affiliation{Laboratory of Mathematical Modeling, Astrakhan State University, Astrakhan, Russia, 414056}

\date{\today}

\begin{abstract}
Numerical simulations using the Monte Carlo method  were performed to study the electrical conductivity of two-dimensional films filled with rodlike particles (rods). The main goal was to investigate the effect of rod alignment on the electrical properties of the films. Both continuous and lattice approaches were used. Intersections of particles were forbidden. Our main findings are (i) both models demonstrate similar behaviors, (ii) at low concentration of rods, both approaches lead to the same dependencies of the electrical conductivity on the concentration of the rods, (iii) the alignment of the rods essentially affects the electrical conductivity, (iv) at some concentrations of partially aligned rods, the films may be conducting only in one direction, (v) the films may simultaneously be both highly transparent and electrically anisotropic.
\end{abstract}

\maketitle

\section{Introduction}\label{sec:intro}

Thin films composed of elongated conductive particles, such as carbon nanotubes, metal nanowires, etc. are of increasing interest, particularly, for the production of flexible transparent conductors (for reviews, see, e.g.,~\cite{Hecht2011AM,McCoul2016AEM,Mutiso2015PPS} and the references therein). Promising applications inspire both experimental studies and simulations of the electrical properties of composite systems with rodlike highly-conducting fillers~\cite{Mutiso2012}.

The transport properties---in particular, the electrical conductivity---of binary systems with conducting fillers inside an insulating host matrix are closely connected with their percolating properties (for reviews, see, e.g.,~\cite{Kirkpatrick1974RMP,Balberg2013book} and the references therein).
One of the first works devoted to the percolation and conductivity of two-dimensional (2D) systems of objects of different shapes, particularly rods, is~\cite{Pike1974PRB}. This seminal work discussed, the behavior of a 2D system of interpenetrating objects of different shapes, in particular rods. Since then the electrical conductivity and percolation phenomena have been extensively simulated for 2D systems using continuous approaches~\cite{Xia1988PRA,Yi2002PRE,Mertens2012PRE,Li2013PRE}.

In recent years, much attention has been paid to the effect of rod alignment on the electrical conductivity, percolation behavior and transparency of thin films.  Of particular interest are  aligned systems based on carbon nanotubes (CNTs)~\cite{Ma2011Carbon}. There are different ways to produce  aligned single-walled carbon nanotubes (for a review, see, e.g.,~\cite{Ma2011Carbon} and the references therein). Particular attention is paid to the effect of the filler alignment on the electrical properties of such composites. The effect of nanotube alignment on percolation conductivity in carbon nanotube/polymer composites has been studied both experimentally and by means of Monte Carlo simulations~\cite{Du2005PRB}. One of the main findings was that the highest conductivity occurs for slightly aligned, rather than isotropic, rods. Monte Carlo simulations have been used to study the effects of nanotube alignment in single-walled carbon nanotube films~\cite{Behnam2007JAP}. These films consist of multiple layers of conductive nanotube networks with percolative transport as the dominant conduction mechanism. The authors reported that minimum resistivity occurred for a partially aligned rather than a perfectly aligned nanotube film. When nanotubes are strongly aligned, the film resistivity becomes highly dependent on the measurement direction~\cite{Behnam2007JAP}. The electrical conductivity of composites with aligned  straight and wavy nanotubes is either lower or higher than that of composites with random nanotube orientation, depending on the degree of alignment; for wavy nanotubes, the highest conductivity occurs when they are slightly aligned~\cite{Li2009JNN}. The type of distribution for a  preferential orientation of CNTs in the network has a drastic effect on the resulting electrical properties~\cite{Simoneau2013JAP}. The relationships between rod alignment, electrical conductivity, percolation behavior and the transparency of thin films have been also discussed in~\cite{Ackermann2015,Ackermann2016}. The electrical conductivity of quasi-2D mono- and polydisperse rod networks having rods of various aspect ratios has been simulated in~\cite{Mutiso2013PRE}.

The critical rod length found in the above work was recalculated  in~\cite{Li2009PRE} as the critical number density $n_c= 5.71 \pm 0.24$ and a more precise value was found $n_c = 5.63726(2)$~\cite{Li2009PRE}. The number density, i.e., number of objects per unit area, is defined as $n = N/L^2$, where $N$ is the total number of objects and $L$ is the linear size of the square region under consideration. This quantity (also denoted as the density, filling factor, or filling density) is the natural quantity used to characterize 2D systems of  widthless rods.

In the effective medium approximation and its modifications, e.g., the so-called generalized effective medium equation, the electrical conductivity of the composite depends on the electrical conductivity of both components (see, e.g., \cite{McLachlan1990JACE}). In methods  involving consideration of the tunneling effect, the conductivity of the matrix is taken into account only by the tunneling between the conductive fillers~\cite{Li2009JNN,Spanos2012PTRSA}. The hopping conductivity in composites made of straight~\cite{Hu2006PRB} and flexible~\cite{Hu2006PRBflexible} metallic nanowires randomly and isotropically suspended in an insulator has been theoretically  studied.

The current distribution in conducting nanowire networks has been studied using analytical as well as Monte-Carlo approaches~\cite{Kumar2017JAP}. The current carrying backbone region  has also been quantified in comparison to isolated and dangling regions as a function of the wire density. The current distribution in the backbone was investigated using Kirchhoff's law.

Recently, the effect of filler alignment on the electrical conductivity of 2D composites has been simulated within a continuous approach for intersecting rodlike particles~\cite{Lebovka2018ARX}.
The multi-scale percolation behavior of the effective conductivity has been studied using a lattice model~\cite{Olchawa2015PhA,Wisniowski2016PhA}. A lattice approach has also  been applied to study the electrical conductivity of a monolayer produced by the random sequential adsorption (RSA) of non-overlapping conductive rodlike particles onto an insulating substrate~\cite{Tarasevich2016}.

In the present research, our investigation is focused on the case of nonintersecting particles both in continuous and lattice approaches. The effects of rod alignment on the electrical conductivity of the films has been compared using both approaches.

The rest of the paper is constructed as follows. In Section~\ref{sec:methods}, the technical details of the simulations are described and all necessary quantities are defined. 
Section~\ref{sec:results} presents our principal findings. Section~\ref{sec:conclusion} summarizes the main results.

\section{Methods}\label{sec:methods}
RSA~\cite{Evans1993} was used to produce a distribution of rods with each desired initial density  and degree of anisotropy. Overlapping with previously deposited rods was strictly forbidden; as a result, a monolayer was formed. Adhesion between deposited rods and the substrate was assumed to be very strong, so once deposited, a rod could not slip over the substrate or leave it (detachment was impossible).

\subsection{Continuous model}\label{subsec:contmodel}
Rods with length $l_s$ and zero thickness, $d_s=0$, (i.e., with infinite aspect ratio, $k=l_s/d_s=\infty$) were randomly and sequentially deposited onto a plane with periodic boundary conditions (PBCs), i.e., onto a torus. Deposition of rods continued until the desired initial number density, $n_0$,  was reached. Basically,  an anisotropic orientation of the  rods was assumed, i.e. the particles were deposited with the given anisotropy. To characterize the anisotropy, we used the mean order parameter calculated as
\begin{equation}\label{eq:S}
 s  = \frac 1 N\sum\limits_{i=1}^N \cos 2 \theta_i ,
\end{equation}
where $\theta_i$ is the angle between the axis of the $i$-th rod and the horizontal axis $x$, and $N$ is the total number of rods in the system (see, e.g,~\cite{Frenkel1985}).

The orientations of rods were distributed according to a normal distribution, i.e., all angles were allowed with different probabilities~\cite{Mietta2014JPhChC}. In this case, the variance of the normal distribution, $\sigma^2$, was connected with the desired mean order parameter as in~\cite{Lebovka2018ARX}
\begin{equation}\label{eq:variance}
  \sigma^2 = -0.5 \ln  s.
\end{equation}
Eq.~\eqref{eq:variance} was used to calculate the variance of the normal distribution providing the desired anisotropy of the system of deposited rods.

Unlike the model in~\cite{Lebovka2018ARX}, a newly deposited particle was not permitted to  overlap previously deposited ones. The kinetics of RSA deposition for such systems has been studied in detail previously~\cite{Sherwood1990,Vigil1990,Ziff1990}. Note that since the rods have zero thickness, jamming is never reached, i.e., the jamming number density $n_j = \infty$. Jamming is the state when no additional object can be placed because all presented voids are too small or their shapes are inappropriate. In contrast, for rods with large but finite aspect ratios $k$, the jamming number density is finite and increases  with $k$ as $n_j\sim k^{0.8}$~\cite{Vigil1989,Ziff1990}.

The length of the system under consideration was $L$ along both the horizontal direction $x$ and the vertical direction $y$. In the present work, all calculations were performed using $L=32 l_s$.

\begin{figure}[htbp]
  \centering
  \includegraphics[width=\linewidth]{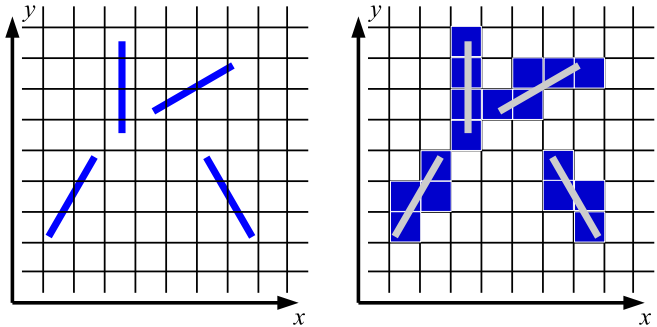}
  \caption{Transformation of continuous system of rods into a discrete system of polyominoes.\label{fig:animals}}
\end{figure}

To calculate the electrical conductivity, a  discretization approach was used.
The plane was covered by a supporting square mesh of size $m\times m$ ($m=64,128,256,512,1024,$ or $2048$). Note that $k^\ast = m/L$ corresponds to the rod length, $l_s$. When a cell of the supporting mesh contains any part of a rod, it is assumed to be occupied and conducting, whilst an empty cell is assumed to be insulating. Discretization transforms the rods into polyominoes of different shapes and sizes, especially, for smaller values of $m$ (see Figure~\ref{fig:animals}). This discretization can produce only one particular kind of polyomino, i.e., those polyominoes which satisfy the condition that a line can be drawn that intersects each cell of the polyomino. In fact, an increase in $m$ at a constant value of $L$  means an increase in the size of the polyomino. In real composites containing CNTs, these fillers, because of their flexibility, are not straight, but have a waved shape. Therefore, modeling with polyominoes presumably better reflects the real situation.

\begin{figure}[htbp]
  \centering
  \includegraphics[width=\linewidth]{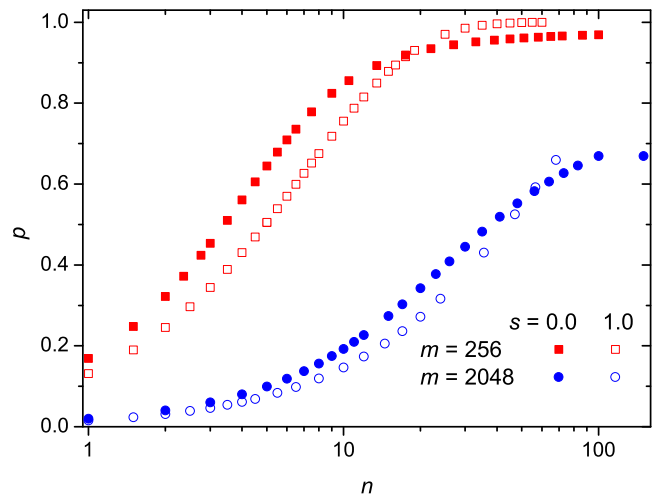}
  \caption{Example of the packing density, $p$, vs. the number density, $n$, for $s = 0, 1$, $m = 256, 2048$. The continuous model. \label{fig:pvsn}}
\end{figure}

To characterize the system after the discretization, the packing density (also denoted as packing fraction) is a convenient quantity. The packing density, $p$, is defined as the number of occupied cells divided by the total number of cells, i.e., $m^2$. There is no one-to-one correspondence between the number density, $n$, and the packing density, $p$, i.e., different placements of the same number of rods at a fixed value of order parameter, $s$, can produce different packing densities after discretization.  The statistical dependence $p(n)$ is close to linear only for a small number density of rods, and it depends on the values of $s$ and $m$ (Figure~\ref{fig:pvsn}). By contrast, for the RSA of objects with non-zero areas (when the jamming concentration is always $p_j<1$), in our continuous model, the real packing density after discretization may reach~1 and, naturally, it remains constant with any further increase in the number of rods. Here, and below, the statistical error is of the order of the marker size.

Forbiddance of intersections of rods obviously does not lead to the forbiddance of intersections of polyominoes. Moreover, polydispersity and polymorphism are inherent properties of the discrete system produced by means of discretization of the continuous system; they never vanish even when $m \to \infty$~\cite{Lebovka2018ARX}. This fact evidences  that the properties of such discrete systems are related but not identical to the properties of the original continuous system.

\subsection{Lattice model}\label{subsec:latticemodel}
In the lattice model, rodlike particles were considered as linear $k$-mers of two mutually perpendicular orientations ($k_x$- and $k_y$-mers) on a square lattice. A linear $k$-mer is a rectangle of size $1 \times  k$ (or $k \times  1$) lattice units, with its corners located at the underlying lattice nodes, i.e., an edge-connected $1 \times  k$ union of cells in the planar square lattice. It may also be defined as a straight polyomino, viz., a straight $k$-omino~\cite{Golomb1954AMM} or $k$-omino of type $I$~\cite{Golomb}.  The deposition of linear $k$-mers onto the 2D square lattice with PBCs (a torus) was performed until a desired packing density, $p \in [0, p_j]$, was reached, where $p_j$ was the jamming packing density. The anisotropic deposition of $k$-mers was examined, i.e., the two  possible orientations of the $k$-mers along the $x$ and $y$ axes had different probabilities.
Since the $k$-mers are allowed to have only two orientations ($\theta = 0$ and $\theta = \pi/2$), the definition of the mean order parameter of the system \eqref{eq:S} reduces to
\begin{equation}\label{eq:s}
s = \frac{N_x - N_y}{N},
\end{equation}
where $N_x$ and $N_y$  are the numbers of $k_x$-mers and $k_y$-mers, respectively, and $N = N_y + N_x$ is the total number of $k$-mers. The value $s = 0$ corresponds to the isotropic system, whereas the value $s = \pm 1$ represents strongly anisotropic alternative (nematic). To ensure deposition with any desired degree of anisotropy, a modification of the RSA has been exploited, viz., so-called relaxation random sequential adsorption (RRSA)~\cite{Lebovka2011}.

\subsection{Computation of the electrical conductivity}

\begin{figure}[htbp]
  \centering
  \includegraphics[width=\linewidth]{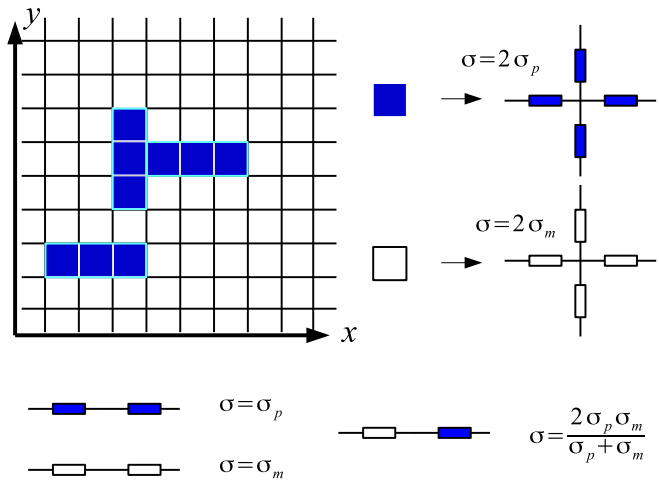}
  \caption{Fragment of a square lattice with three deposited 3-mers of different orientations. All possible combinations of the conductivities are indicated.\label{fig:transformation}}
\end{figure}

To transform the lattice into a random resistor network (RRN), the PBCs were removed (the torus was unwrapped into a plane) and each cell was associated with a set of 4 conductors. Different electrical conductivities corresponding to the empty cells, $\sigma_m$, occupied cells, $\sigma_p$,  and between empty and occupied cells, $\sigma_{pm}=2\sigma_p \sigma_m / (\sigma_p+\sigma_m)$ were assumed (Figure~\ref{fig:transformation}). A large contrast in electrical conductivity was assumed,  $\Delta=\sigma_p/\sigma_m \gg 1$. We put $\sigma_m =1$, $\sigma_p = 10^6$ in arbitrary units.

In our calculations,  two conducting buses subjected to a potential difference were applied to the opposite borders of such a plane.  Electrical conductivity was calculated in the direction along the alignment of the particles (longitudinal conductivity, $\parallel$)  and in the direction across the alignment (transversal conductivity, $\perp$)  (see~\cite{Lebovka2016,Lebovka2017} for details).

Two algorithms were  applied to calculate the electrical conductivity, viz., the Frank--Lobb algorithm~\cite{Frank1988} (continuous model) and the direct electrifying algorithm~\cite{Li2004JPhysA} (lattice model).
For a quantitative description of the anisotropy of the electrical conductivity in the $x$ and $y$ directions, the anisotropy ratio, $\delta$, as defined from the electrical contrast, $\Delta$,
\begin{equation}\label{eq:deltaxy}
\sigma_\parallel /\sigma_\perp =\Delta^\delta,
\end{equation}
was used~\cite{Tarasevich2016}. $\delta=0$ for isotropic systems and $\delta\approx 1$ for highly anisotropic systems with
$\sigma_\parallel /\sigma_\perp \approx \Delta$.

To characterize the insulator--conductor transition, we used the value
\begin{equation}\label{eq:sigmag}
  \sigma_g =\sqrt{\sigma_m \sigma_p}.
\end{equation}
We treated a system with conductivity $\sigma > \sigma_g$ as conducting while a system with conductivity $\sigma < \sigma_g$ was considered as insulating. We denoted values of the number density and the packing density corresponding to $\sigma_g$ as $n_g$ and $p_g$, respectively. Basically, a lattice size of $L = 100k$ was used and all the quantities under consideration were averaged over 10 independent statistical runs, unless otherwise explicitly specified in the text.

Using the continuous model, the effects on the electrical conductivity of the packing density of rods, $p$, and the anisotropy in their orientation, $s$, as well as the size of the supporting square mesh, $m$, were investigated.

With the lattice model, we studied the effect of $k$-mer length and the anisotropy of their deposition on the electrical conductivity, $\sigma$,  of the monolayer. The values of $k$ were $2, 4, 8,\dots, 128$.

\section{Results}\label{sec:results}
\subsection{Continuous model\label{subsec:contres}}

\begin{figure}[htbp]
  \centering
\includegraphics[width=\linewidth]{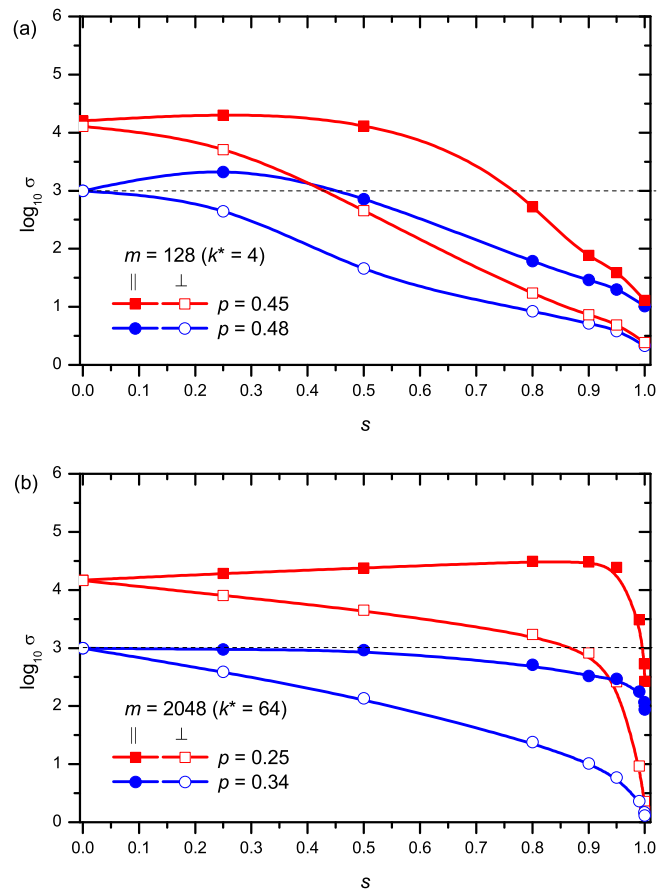}
  \caption{Continuous model: Longitudinal and transversal effective electrical conductivity, $\sigma$, vs. order parameter, $s$, for (a) $m = 128$, $n = 1.4, 1.5$ and (b) $m = 2048$, $n = 13.56, 20$. The dashed line corresponds to the value $\sigma_g$. The solid lines are provided simply as visual guides. \label{fig:cond2048s}}
\end{figure}

Figure~\ref{fig:cond2048s} demonstrates the dependencies of the longitudinal and transversal effective electrical conductivities, $\sigma$,  versus the order parameter, $s$, for $m = 128$ ($k^\ast = 4$), number density $n = 1.4, 1.5$ ($p = 0.45, 0.48$) and $m = 2048$ ($k^\ast=64$), $n = 13.56, 20$ ($p = 0.25, 0.34$). The smaller values correspond to $\sigma(0) = \sigma_g$. A  system, which in the isotropic state ($s = 0$) was at the insulator--conductor transition, becomes an insulator in both directions when the order parameter approaches 1.

\begin{figure}[htbp]
  \centering
  \includegraphics[width=\linewidth]{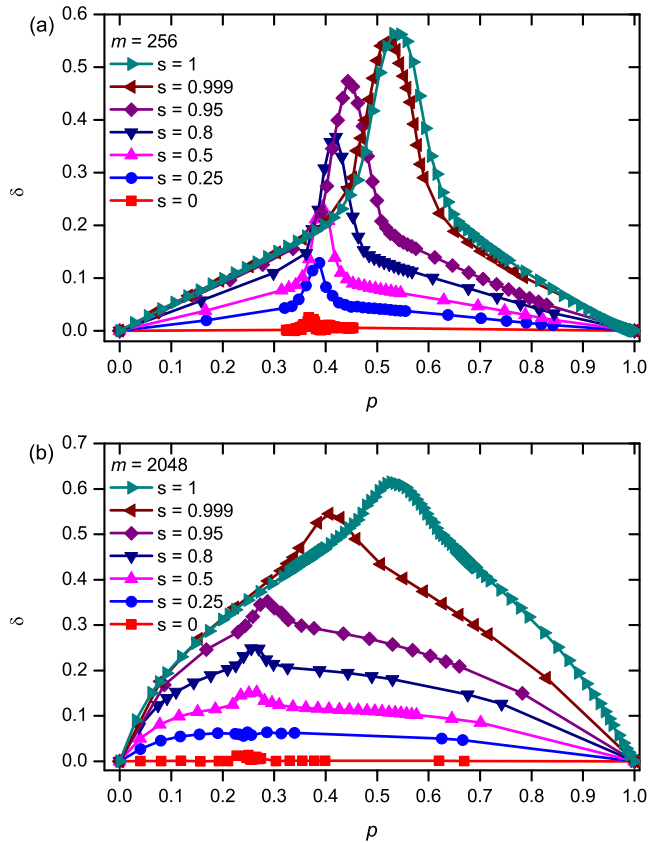}
  \caption{Continuous model: Anisotropy of the effective electrical conductivity, $\delta$ (Eq.~\ref{eq:deltaxy}),  vs. packing density, $p$, for different values of the order parameter, $s$. (a) $m= 256$ ($k^\ast = 8$), (b) $m = 2048$ ($k^\ast = 64$). The results are averaged over 10 independent statistical runs, except at $s = 1$ (5 runs). The lines are provided simply as visual guides.\label{fig:m2048delta}}
\end{figure}

Figure~\ref{fig:m2048delta} presents examples of the anisotropy of the effective electrical conductivities, $\delta$, versus the packing density $p$, for different values of the order parameter $s$ and for $m = 2048$ and $m = 256$. With an increase in the order parameter, $s$, the anisotropy increases. The position of the maximum on the curves shifts toward a larger value of $p$ as the value of $s$ increases. The plots show curvatures of different signs outside the vicinity of the maximum for $m = 2048$ and $m = 256$. Figure~\ref{fig:m2048delta} suggests that a sample may have high electrical anisotropy and be transparent when the fillers are long enough and are not perfectly aligned in one direction.

\begin{figure}[htbp]
  \centering
  \includegraphics[width=\linewidth]{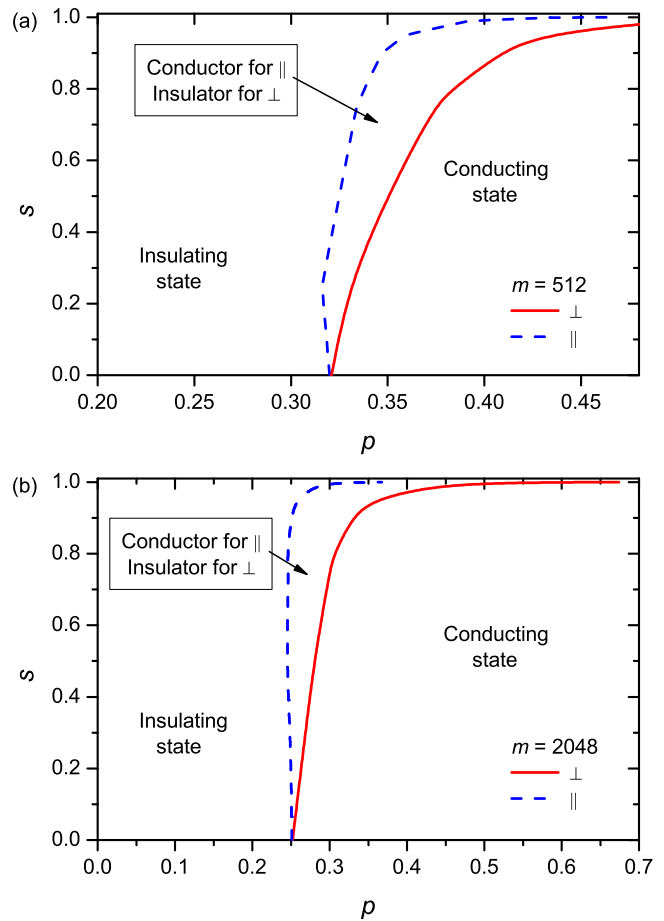}
  \caption{Continuous model: Examples of the phase diagram in the $(p, s)$-plane. (a) $m=512$ ($k^\ast = 16$),  (b) $m=2048$ ($k^\ast = 64$).
  \label{fig:phasecont}}
\end{figure}

This insulator--conductor transition ($\sigma(p,s)=\sigma_g$) occurs at different values of the order parameter, $s$, depending on the packing density, $p$. Hence, for each given value of $k$, there is the phase diagram in a plane $(p,s)$ (Figure~\ref{fig:phasecont}). The critical curves $s(p)$  divide the phase plane $(p,s)$ into two regions, viz., conducting and insulating. The region between the two curves corresponds to those samples which are conductors in one direction and insulators in the perpendicular direction. Note that these  curves are rather nominal, since the phase transition is not sharp but very diffuse.

\subsection{Lattice model}\label{subsec:latticeres}

Primary attention has been paid to the particular packing density $p^\ast \approx 0.5(p_c(1) + p_c(0))$, where $p_c(s)$ is the percolation threshold for the particular value of the mean order parameter $s$ (see Table~\ref{tab:pc}). This choice is determined by the fact that, at this packing density, the system is expected to undergo the conductor--insulator phase transition when the order parameter changes from 0 to 1.
\begin{table}[!htbp]
  \caption{Percolation threshold $p_c$  for $k$-mers of different length $k$ and two particular values of the order parameter $s=0$ and $s = 1$, $L \to \infty$ (extracted from~\cite{Tarasevich2012}).}\label{tab:pc}
\begin{ruledtabular}
  \begin{tabular}{cccccccc}
    $s\,\backslash\, k$ &  2  &  4  &  8  &  16 &  32 &  64 &  128 \\
\hline
0.0 &  0.5619 &  0.5050 &  0.4697 &  0.4638 &  0.4748 & 0.4928 & 0.5115\\
1.0 &  0.5862 &  0.5672 &  0.5526 &  0.5442 &  0.5397 & 0.5376 & 0.5366  \\
  \end{tabular}
  \end{ruledtabular}
\end{table}

\begin{figure}[htbp]
  \centering
  \includegraphics[width=\linewidth]{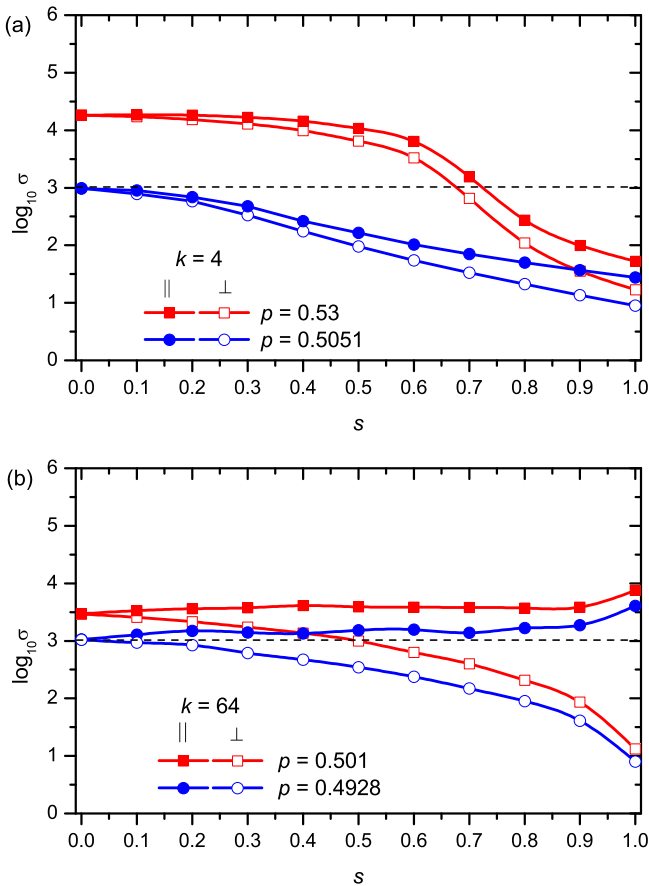}
    \caption{lattice model: Examples of the effective longitudinal and transversal electrical conductivities  vs.  order parameter, $s$, for two particular values of the packing density, $p=p^\ast$ (boxes) and $p=p_c$ (circles). (a) $k = 4$, $L=10^3k$. (b) $k=64$, $L=100k$. The dashed line corresponds to the value $\sigma_g$. The solid lines are provided simply as visual guides. \label{fig:condk4k64}}
\end{figure}

Figure~\ref{fig:condk4k64} shows examples of the dependencies of the electrical conductivity, $\sigma$, on the order parameter, $s$, for $k = 4$ and $k = 64$. The data are presented for two particular values of the packing density, $p = p_c (0)$ and $p=p^\ast$, where $p_c(s)$. Here, the value of $p=p^\ast$ corresponds to the mean percolation packing density for systems with order parameters $s=0$ and $s=1$ and this value can be useful for analysis of the conductivity behavior with changes of the order parameter, $s$.

Clear anisotropy of the electrical conductivity was observed at both $p = p_c (0)$ and $p=p^*$. At $p = p_c (0)$ for short rods ($k=4$, Figure~\ref{fig:condk4k64}(a)) where an increase in the order parameter $s$ resulted in a decrease of  the values of both the longitudinal and transversal conductivities as a result the system going into an insulating state. However, for long rods ($k=64$, (Figure~\ref{fig:condk4k64}(b)) different behavior was observed, viz., with an increase of $s$ the system remained in the conducting state for the longitudinal direction whereas it went into the insulating state for the transversal direction.

The conductor--insulator phase transition was clearly seen when the packing density was close to the value $p^\ast$ (Figure~\ref{fig:condk4k64}). For small values of $k$ ($k \lessapprox 16$), this transition in the longitudinal  and transversal conductivities one occurs over a small range of values of the order parameter, $s$ (Figure~\ref{fig:condk4k64}(a)). With increasing $k$, the values of $s$ for the transition in the two directions are increasingly different. For the value $k = 32$, the longitudinal conductor--insulator transition no longer occurs, the lattice remaining a conductor for all values of $s$. Monolayers produced by deposition of long rods essentially demonstrate electrical anisotropy, such they may be conducting in one direction and insulating in the perpendicular direction. The example for $k=64$ is presented in Figure~\ref{fig:condk4k64}(b). At $p \approx p_c$, the system has similar electrical properties along both directions, $\sigma \approx 10^3$ when the order parameter $s=0$. When the order parameter $s$ increases from 0 to 1, the transversal electrical conductivity decreases, whereas  the longitudinal conductivity increases. Thus,  when $s=1$, anisotropy of the placement of the fillers  produces anisotropy in the electrical conductivity.

\begin{figure}[htbp]
  \centering
  \includegraphics[width=\linewidth]{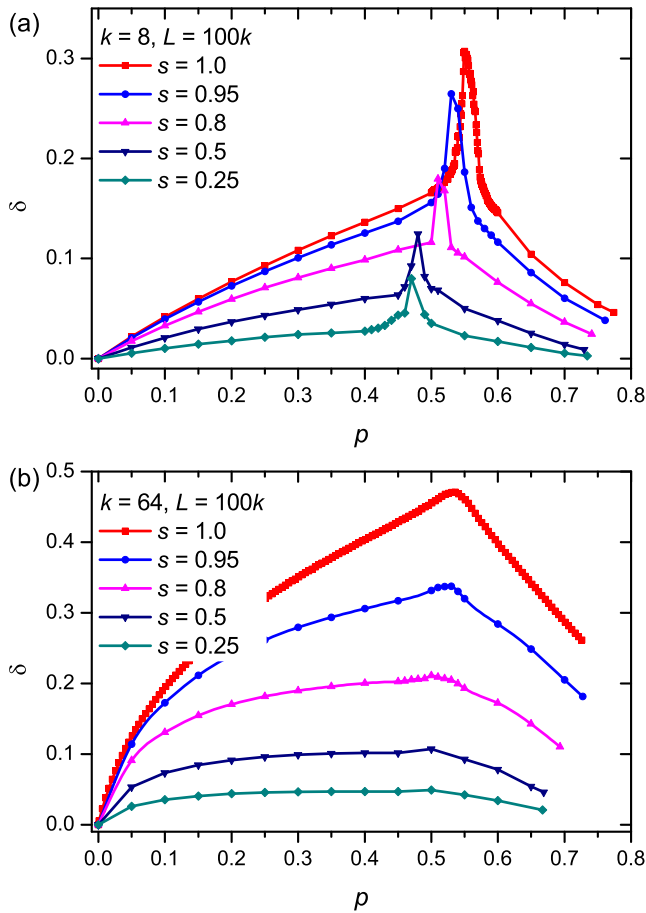}
    \caption{Lattice model: Anisotropy of the effective electrical conductivity, $\delta$ (Eq.~\ref{eq:deltaxy}),  vs. the packing density, $p$, for different values of the order parameter, $s$. (a) $k= 8$, (b) $k = 64$. The results are averaged over 10 independent statistical runs. The lines are provided simply as visual guides.\label{fig:delta}}
\end{figure}

An increase in the anisotropy of the electrical properties, $\delta$, with an increase in the order parameter, $s$, is shown in Figure~\ref{fig:delta} for different values of $k$. The larger the values of $k$ and $s$ the more significant the increase in the anisotropy that can be observed.

\begin{figure}[htbp]
  \centering
  \includegraphics[width=\linewidth]{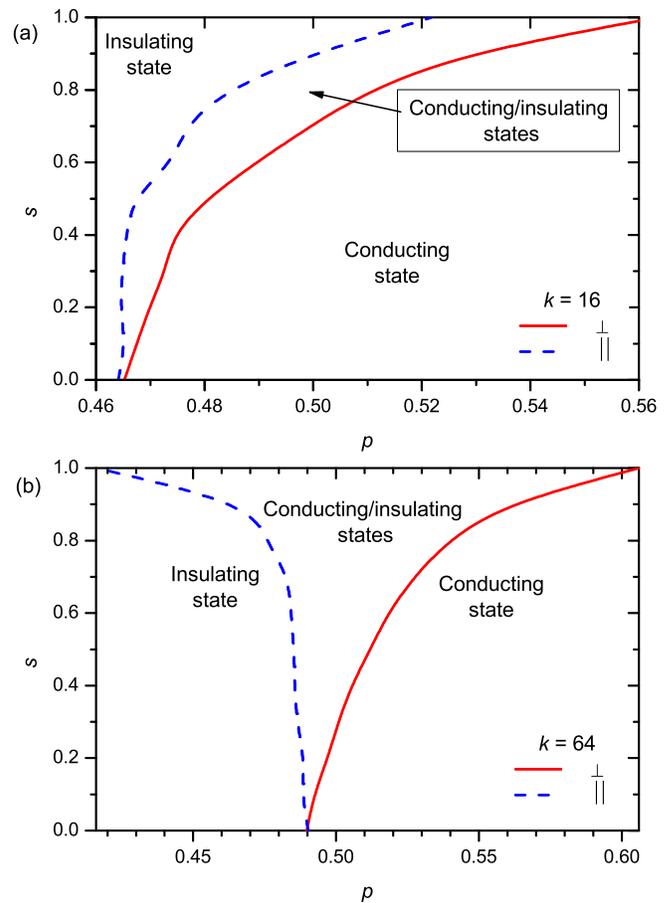}
  \caption{Lattice model: Examples of the phase diagrams in the $(p, s)$-plane. (a) $k = 16$,  (b) $k = 64$.  $L=100k$.\label{fig:phasek16}}
\end{figure}

It is of note that the phase diagrams for $k=16$ and $k=64$ are quite different (Figure~\ref{fig:phasek16}). Namely, for $k=16$, the system undergoes the transition from an insulating to a conducting state through a highly anisotropic state both at a fixed value of packing density when the order parameter increases and at a fixed value of the order parameter when the packing density increases.  In contrast, for $k=64$, increase in the order parameter never leads to a highly anisotropic state at any value of the packing density, the system bypassing the highly anisotropic state as it undergoes the transition from a conducting to an insulating state. However, a highly anisotropic state is reachable during the transition from insulator to conductor when the order parameter is fixed while the packing density increases.

\subsection{Comparison of the models}

\begin{figure}[htbp]
  \centering
  \includegraphics[width=\linewidth]{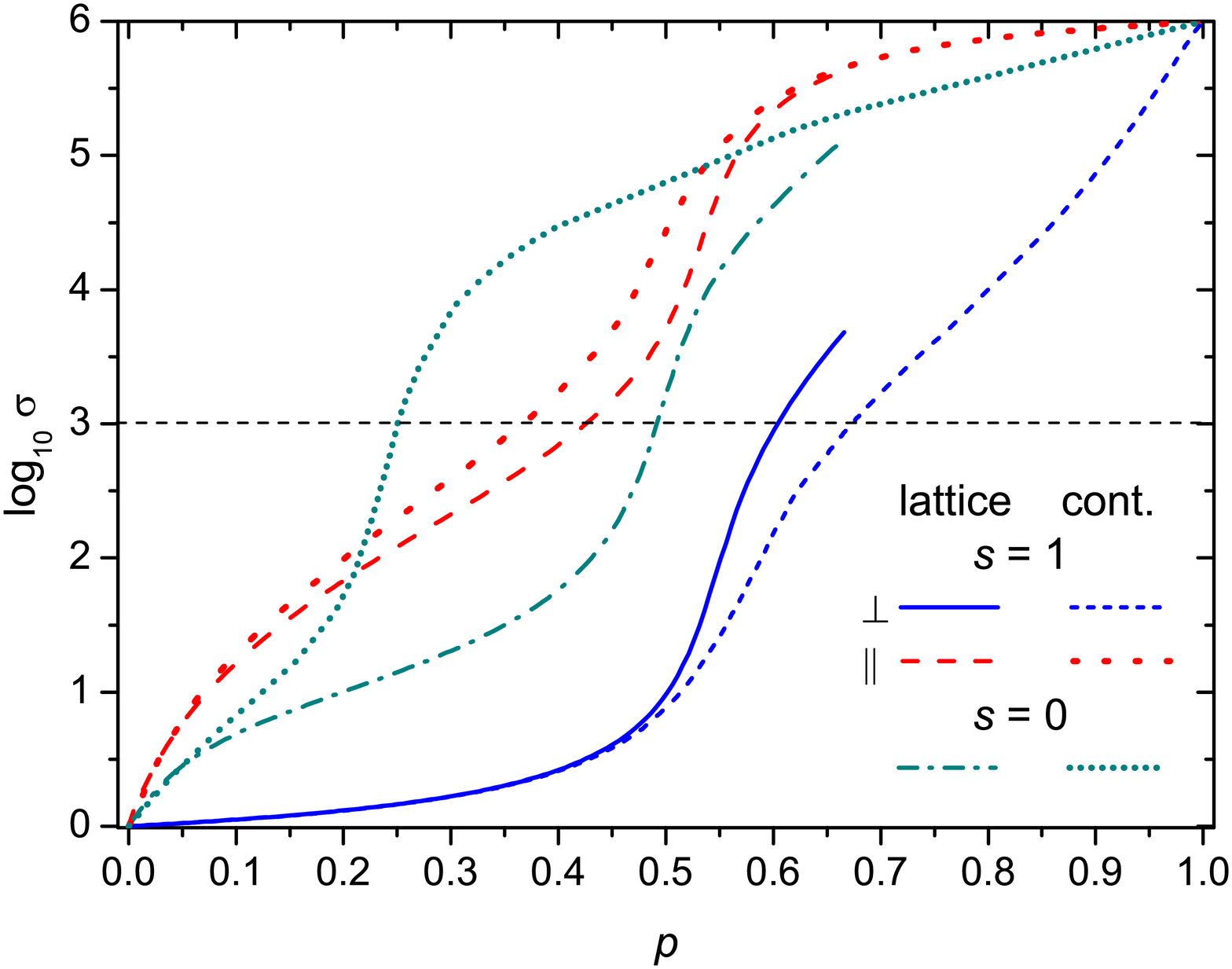}
  \caption{Example of the longitudinal and transversal effective electrical conductivities  vs.  packing density $p$, for values of the order parameter $s = 0, 1$. For the continuous model, $m = 2048$ ($k^\ast=64$), the results are averaged over 10 independent statistical runs, except at $s = 1$  (5 runs). For the lattice model, $k = 64$,  $L = 100k$, the results are averaged over 10 independent statistical runs.  The dashed line corresponds to the value $\sigma_g$. \label{fig:Condpm2048k64}}
\end{figure}

Figure~\ref{fig:Condpm2048k64} compares the dependencies of the longitudinal and transversal effective electrical conductivities, $\sigma$, versus the packing density, $p$, with $s = 0, 1$, obtained using each of the continuous and the lattice models. For the continuous model $m = 2048$ ($k^\ast =64$), and for the lattice model $k = 64$, $L = 100k$.  Note that the quantity
\begin{equation}\label{eq:ic}
[\sigma] = \left.\frac{\mathrm{d}\ln\sigma}{\mathrm{d}p}\right|_{p \to 0},
\end{equation}
is called the ``intrinsic conductivity'' (see, e.g.,~\cite{Douglas1995AdvChPh,Garboczi1996PRE}). The ``intrinsic conductivity'' is equal to the slope of the tangent to the curve $\sigma(p)$ at the point $p=0$. It is remarkable, that both models demonstrate indistinguishable dependencies of their electrical conductivities on the packing density, when $p \approx 0$, i.e. the intrinsic conductivities calculated with both the continuous and lattice models are the same. Moreover, both the continuous and the lattice models show similar behavior of the dependencies of the electrical conductivities on the packing density when $s=1$. In the anisotropic case, $s = 1$, no large differences in the results from the two models can be observed. When all the particles are oriented  in one direction, the rods deposited onto a plane are closely analogous to  $k$-mers deposited onto a square lattice. In the isotropic case, the differences are fairly noticeable. For $s=0$, the conductor--insulator transition in the case of the continuous model occurs at a smaller value of packing density, $p$. These differences are apparently due in part to the fact that, in the lattice model, the $k$-mers are equiprobably oriented along only two mutually perpendicular directions, whereas, in the continuous model, the rods are placed equiprobably in all directions. In the latter case, the rods generate polyominoes of different shapes and sizes, and this seems to lead to the appearance of additional paths for the electrical current. Although the size of the $k$-mers chosen for comparison is equal to the characteristic length of the rods after discretization,  $k^\ast = m/L =2048/32 = 64$, this correspondence on the basis of size is somewhat nominal for the isotropic case.

Nevertheless, the models demonstrate quite different properties in the ordered (nematic) state ($s=1$) when the packing density, $p$, corresponds to $\sigma(0) = \sigma_g$. In the continuous model, the film is insulating in both directions, while, in the lattice model, the film is insulating in one direction and conducting in the perpendicular direction (compare Figures~ \ref{fig:cond2048s} and \ref{fig:condk4k64}).

For both models, the anisotropies of the effective electrical conductivities demonstrate similar behavior (Figures~\ref{fig:delta} and \ref{fig:m2048delta}). Nevertheless, the quantitative differences are quite remarkable, viz., the anisotropy is approximately two-fold  lower in the lattice model than in the continuous model.

\section{Conclusion}\label{sec:conclusion}
The effect of the alignment of rodlike particles on the electrical conductivity
of 2D composites has been investigated using continuous and lattice models. In both models, highly conductive elongated particles were randomly placed, without intersections, on a poorly conductive substrate. In the lattice model, the $k$-mers were constrained to only the horizontal and vertical directions, while, in the continuous model, the zero-width rods could be placed with any planar  orientation. To calculate the electrical conductivity of the layer in the continuous model, the layer was further discretized, which made it possible to reduce the situation to a problem of the effective sizes of the resulting polyominoes on the substrate. Our studies have shown that, as well as size and concentration of the fillers, their alignment affects the electrical conductivity of the monolayer. Since both models demonstrated similar behavior, the use of the lattice model for  qualitatively explaining the electrical properties of composite materials containing elongated objects looks quite reasonable. Our simulations suggest that highly transparent and electrically anisotropic 2D composites can be produced by the deposition of almost-aligned elongated conductive particles onto a transparent insulating substrate.

In both models, the dependencies of the electrical conductivities on the packing densities are the same when the packing density is small. This means that the intrinsic conductivities in the case of the continuous model are exactly the same as previously calculated within the framework of the lattice model~\cite{Tarasevich2016}.

The continuous model has an obvious drawback, viz., any cell of the supporting mesh will be treated as conductive without respect to the number of rods intersecting the cell, their orientations or their positions inside such a cell. The obvious way to improve the model is accounting such effects. Nevertheless, such an enhancement is unlikely to change the qualitative behavior of the system, although  the quantitative changes would be very likely.

\begin{acknowledgments}
The authors acknowledge funding provided by the Ministry of Education and Science of the Russian Federation, Project No.~3.959.2017/4.6 (Y.Y.T., I.V.V., V.A.G., V.V.C., and A.V.E.) and the National Academy of Sciences of Ukraine, Project No.~43/18-H and 15F (0117U006352)  (N.I.L.). We would like to thank V.V.~Laptev for the programs that had been previously developed for other purposes but were also used in the research presented here.
\end{acknowledgments}

\bibliography{sticks,conductivity,Drying,alignedrods,polyominoes}

\end{document}